\documentclass[12pt]{article}

\usepackage{epsfig}
\usepackage{amssymb}
\usepackage{amsmath}
\usepackage{amsfonts}
\usepackage{graphicx}
\usepackage{mathrsfs}
\usepackage[dvips]{color}
\usepackage{multirow}

\numberwithin{equation}{section}

\newcommand{\fu}{\mathfrak{u}}

\newcommand{\fU}{\mathfrak{U}}

\newcommand{\bD}{\mathbf{D}}

\newcommand{\bS}{\mathbf{S}}

\newcommand{\cD}{\mathcal{D}}

\newcommand{\cG}{\mathcal{G}}

\newcommand{\cO}{\mathcal{O}}

\newcommand{\cR}{\mathcal{R}}
\newcommand{\cS}{\mathcal{S}}

\newcommand{\cZ}{\mathcal{Z}}
\newcommand{\be}{\begin{equation}}
\newcommand{\ee}{\end{equation}}
\newcommand{\bea}{\begin{eqnarray}}
\newcommand{\eea}{\end{eqnarray}}
\newcommand{\nn}{\nonumber}

\newcommand{\ed}{\end{document}}

\newcommand{\bi}{\begin{itemize}}
\newcommand{\ei}{\end{itemize}}

\newcommand{\bce}{\begin{center}}
\newcommand{\ece}{\end{center}}

\newcommand{\p}{\partial}
\newcommand{\psl}{p\!\!\!/}

\begin{document}

\title{Fermionic one-particle states in curved spacetimes}

\author{Farhang Loran\thanks{E-mail address: loran@cc.iut.ac.ir} \\[6pt] Department of Physics, Isfahan University of Technology, \\ Isfahan
84156-83111, Iran\\[6pt] }

\date{ } \maketitle

\begin{abstract}
We show that a notion of one-particle state and the corresponding vacuum state exists in general curved backgrounds for  spin $\frac{1}{2}$ fields.
 A curved spacetime can be equipped with a coordinate system in which the metric component $g_{--}=0$. We separate the component of the left-handed massless Dirac field which is annihilated by the null vector  $\partial_-$ and compute the corresponding Feynman propagator. We find that the  propagating modes are localized on  two dimensional subspaces and  the Feynman propagator  is similar to the Feynman propagator of chiral fermions in two dimensional Minkowski spacetime. Therefore, it can be interpreted in terms of one-particle states and the corresponding vacuum state similarly to the second quantization in Minkowski spacetime.
 \end{abstract}



\maketitle

\section{Introduction}
 The study of massless Dirac fermions on non-stationary curved backgrounds has diverse theoretical and observational motivations \cite{Parker71,Bunch1979,Anber:2016yqr,Cortez:2016xsn,Eichhorn:2016vvy,Singh:2008sx}.   Although the concepts of the vacuum state and the one-particle states are at the core of the second quantization in Minkowski spacetime, in general, they are not well-defined  in quantum field theory in time-dependent curved spacetimes \cite{Birrel-book}.

 In principle,  the vacuum state can be inferred from a plausible two-point function of the quantum fields \cite{Fulling-book}.  Such a two-point function can be computed by path integrals.  For Dirac spinors it is given by
    \be
    S_F(x,x')=\cZ^{-1}\int D\overline\psi D\psi e^{i\cS}\psi(x)\overline\psi(x'),
    \label{1st}
    \ee
where $\overline\psi$ and  $\psi$ are  Grassmann fields, $\cS$ denotes the action and
    \be
    \cZ:=\int D\overline\psi D\psi e^{i\cS}.
    \label{partition-function}
    \ee
Eq.\eqref{1st}  implies that  $S_F(x,x')$ is a  Green's function for  Dirac operator, i.e., it solves Dirac equation with the Dirac delta-function source, satisfying certain boundary conditions.

In four dimensional Minkowski spacetime,  Dirac operator in an inertial (non-accelerating) reference frame is $(i\partial\!\!\!/-m)$,\footnote{$m$ denotes the mass of Dirac particle,  $\partial\!\!\!/:=\gamma^a\partial_a$ and $\gamma^a$ are Dirac matrices.} and $S_F(x,x')$  equals the Feynman propagator\footnote{ $\theta$ denotes the Heaviside step function. $\theta(x)$ equals 1 and 0, for $x>0$ and $x<0$ respectively.}
    \be
    S_F(x,x')=S^A(x,x')\theta(x^0-{x'}^0)-S^B(x,x')\theta({x'}^0-x^0),
   \label{2nd}
    \ee
in which, the amplitudes
    \begin{align}
    &S^A(x,x'):=\left\langle\psi(x)\Big|\overline\psi(x')\right\rangle,\\
    &S^B(x,x'):=\left\langle\overline\psi(x')\Big|\psi(x)\right\rangle,
     \end{align}
represent the propagation of {\em positive-energy} particles from $x'$ to $x$ and from $x$ to $x'$ respectively. That is, we suppose that\footnote{$E_p:=\sqrt{{\vec p}^2+m^2}$. $p\cdot x:=\eta_{ab}p^ap^b$ where $p^0:=E_p$, and the Minkowski metric $\eta_{ab}={\rm diag}(1,-1,-1,-1)$.}
    \begin{align}
    \label{psi-R}
    &\left|\psi(x)\right\rangle=\sum_s\int\frac{d^3p}{(2\pi)^3\sqrt{2E_p}} v^s(p)e^{ip\cdot x}\left|\vec p,s;B\right\rangle,\\
    \label{psi-L}
    &\left\langle\psi(x)\right|=\sum_s\int\frac{d^3p}{(2\pi)^3\sqrt{2E_p}} u^s(p)e^{-ip\cdot x}\left\langle\vec p,s;A\right|,
    \end{align}
where $v(p)e^{ip\cdot x}$ and $u(p)e^{-ip\cdot x}$ are solutions of Dirac field equation,
    \begin{align}
    \label{v-eom}
    &(\psl+m)v(p)=0,\\
    \label{u-eom}
    &(\psl-m)u(p)=0,
    \end{align}
with positive energy and negative energy respectively, and $A$ and $B$ denote other quantum numbers collectively. For example, for fermions coupled to Maxwell field, $A$ and $B$ denote the electric charges $q$ and $-q$ respectively.
The expressions for $\left|\overline\psi(x)\right\rangle$ and $\left\langle\overline\psi(x)\right|$ follows the definition $\overline\psi:=\psi^\dag\gamma^0$. Eqs.\eqref{psi-R} and \eqref{psi-L} give, respectively,
\begin{align}
    \label{psibar-L}
    &\left\langle\overline\psi(x)\right|=\sum_s\int\frac{d^3p}{(2\pi)^3\sqrt{2E_p}} {\bar v}^s(p)e^{-ip\cdot x}\left\langle\vec p,s;B\right|,\\
    \label{psibar-R}
    &\left|\overline\psi(x)\right\rangle=\sum_s\int\frac{d^3p}{(2\pi)^3\sqrt{2E_p}} {\bar u}^s(p)e^{ip\cdot x}\left|\vec p,s;A\right\rangle.
    \end{align}

Lorentz transformation and the  field equations \eqref{v-eom} and \eqref{u-eom} give
    \begin{align}
    \label{Int-v-completeness}
    &\sum_s v^s(p){\bar v}^s(p)=\psl-m,\\
    \label{Int-u-completeness}
    &\sum_s u^s(p){\bar u}^s(p)=\psl+m,
    \end{align}
where we have also fixed  the normalization constants.
Translational and rotational invariance imply that\footnote{$\delta^D$ denotes  the Dirac delta-function in $D$-dimensions. The Kronecker delta $\delta^{rs}$ equals 1 and 0 for $r=s$ and $r\neq s$ respectively.}  \cite{Peskin-book}
    \begin{align}
    \label{orthogonal-A}
    &\left\langle \vec p,s;A\Big|\vec q,r;A\right\rangle=(2\pi)^3\delta^3(\vec p-\vec q)\delta^{rs},\\
    \label{orthogonal-B}
    &\left\langle \vec p,s;B\Big|\vec q,r;B\right\rangle=(2\pi)^3\delta^3(\vec p-\vec q)\delta^{rs}.
    \end{align}
Consequently,
    \begin{align}
    \label{g1}
    &S^A(x,x')=\int\frac{d^3p}{(2\pi)^3}\frac{\psl+m}{2E_p}e^{-ip\cdot(x-x')},\\
    \label{g2}
    &S^B(x,x')=\int\frac{d^3p}{(2\pi)^3}\frac{\psl-m}{2E_p}e^{ip\cdot(x-x')}.
    \end{align}

The spinors anti-commute at spacelike separation, i.e.,
  \begin{align}
  &S^A(x,x')=-S^B(x,x'),&(x-x')^2<0.
  \label{Fermi-spacelike}
  \end{align}
This in turn indicates that the particles represented by Dirac field obey  Fermi statistics  \cite{Peskin-book}. That is, if we postulate  a  {\em vacuum state} $\left|0\right>$ invariant under translation and Lorentz transformation, and  {\em ladder operators} ${a^r_p}$ and ${b^r_p}$ such that
    \begin{align}
    &{a^r_p}\left|0\right>=0,&{b^r_p}\left|0\right>=0,\\
    &\left|\vec p,r;A\right\rangle={a^r_p}^\dag\left|0\right>,    &\left|\vec p,r;B\right\rangle={b^r_p}^\dag\left|0\right>,
    \end{align}
we obtain
    \begin{align}
    &{a_p^r}^\dag{a_q^s}^\dag\left|0\right\rangle=-{a_q^s}^\dag {a_p^r}^\dag\left|0\right\rangle,\\
    &{b_p^r}^\dag{b_q^s}^\dag\left|0\right\rangle=-{b_q^s}^\dag {b_p^r}^\dag\left|0\right\rangle.
    \end{align}

In general,  the concept of {\em positive energy} is lost in non-stationary  curved spacetime and  also  in non-inertial reference frames in flat spacetime\footnote{A counterexample is the non-inertial frame corresponding to Rindler observers in Minkowski spacetime  because the Rindler wedge of Minkowski spacetime is globally hyperbolic \cite{Fulling73}.} \cite{Wald-Book}. As we have seen, this concept is  essential for defining  {\em Dirac particles}   in inertial reference frames in Minkowski spacetime. Related to this fact, we do not have, in general, a well-defined concept of time-ordering which has been used in Eq.\eqref{2nd}, although the Feynman propagator $S_F(x,x')$ itself is well-defined through Eq.\eqref{1st}.

In principle, physical observables in accelerating frames in Minkowski spacetime can be described in terms of their counterparts in inertial frames by applying the corresponding  local Lorentz transformations.\footnote{There are also nonlocal effects such as the appearance of event horizons in the accelerating frames,  that add nontrivial features to such descriptions.  For example a uniformly accelerated observer in Minkowski spacetime has access only to a subset of physical states which correspond to the  Rindler wedge of  the Minkowski spacetime \cite{Susskind-book}.}

In this paper we study massless spinors on curved spacetimes. In general, a four dimensional curved spacetime can be equipped with a coordinate system $(x^\pm,{\bf x}_\bot)$,  where
    \begin{align}
    &x^\pm:=x^0\pm x^3,\\
    &{\bf x}_\bot:=(x^1,x^2),
    \end{align}
 such that  $g_{--}=0$. In a frame of reference given by\footnote{$\nabla_\mu$ denotes the Levi-Civita connection and $g$ is the determinant of the spacetime metric. ${e_\mu}^{(a)}$  is the tetrad identifying the frame of reference. $\partial_{(a)}:={e^\mu}_{(a)}\partial_\mu.$}
     \begin{align}
     &\partial_{(-)}=(-g)^{-\frac{1}{2}}\partial_-,\\
     &{e^\nu}_{(1)}\nabla_- e_{\nu(2)}=0,
     \end{align}
we separate the spin-up component of the left-handed massless Dirac field along the third direction  and compute the corresponding Feynman propagator $S_F^{\uparrow(L)}(x,x')$. We find that
    \be
    S_F^{\uparrow(L)}(x,x')=S^{+}_F(x,x')\delta^2({\bf x}_\bot-{\bf x}_\bot'),
    \label{Int-SF-4D}
    \ee
 where $S^{+}_F(x,x')$ denotes the Feynman propagator of left-moving massless fermions  in an inertial frame in a two dimensional Minkowski spacetime equipped with coordinates $(x^0,x^3)$,\footnote{${\rm sgn}(x):=\theta(x)-\theta(-x)$ is the sign function.}
    \be
    S_F^{+}(x,x')=\frac{1}{2}{\rm sgn}(x^0-{x'}^0)\delta(x^+-{x'}^+)-\frac{i}{2\pi}\int_0^\infty dp\,\sin\! \left[p(x^+-{x'}^+)\right].
    \label{Int-SF-2D}
    \ee
 Therefore, $S_F^{\uparrow(L)}(x,x')$  can be interpreted in terms of one-particle states and the corresponding vacuum state, similarly to the second quantization in two dimensional  Minkowski spacetime.  Eq.\eqref{Int-SF-4D}  shows that the corresponding modes are localized on the ${\bf x}_\bot$ plane indeed. The sign function in Eq.\eqref{Int-SF-2D} can be equivalently  given in terms of $x^-$ because
    \be
    {\rm sgn}(x^0-{x'}^0)\delta(x^+-{x'}^+)={\rm sgn}(x^--{x'}^-)\delta(x^+-{x'}^+).
    \label{Int-switch-ordering}
    \ee
 Since $x^-$ is singled out by its definition via $g_{--}=0$, this substitution leads to a covariant description of the  Feynman propagator.

Our paper is organized as follows. In section \ref{section-general-prop}, we review the general properties of Feynman propagator given  by the path integral \eqref{1st}. Section \ref{section-flat} is devoted to review the second quantization of Dirac field in inertial frames in two dimensional flat spacetime. Our goal in section \ref{section-upside-down} is to explain how the concepts of  vacuum state and  one-particle states can be inferred from  the short distance singularity of the  Feynman propagator.

In section \ref{section-fermion}, we compute the Feynman propagator  in a two dimensional curved background. Obviously, the result depends on the frame of reference. In section \ref{section-frame}, we introduce the left-handed frame in which the  curvature effect is removed from the left-mover sector and the Feynman propagator is given by Eq.\eqref{Int-SF-2D}. Following our discussion in section \ref{section-upside-down}, we conclude that even in a non-stationary spacetime, left-mover vacuum state and  left-mover one-particle states are well-defined in the left-handed frame.  Of course, by removing the curvature effect  from the left-handed sector,  the whole effect of the spacetime curvature transmits to the right-handed sector, thus we do not expect a universally consistent definition of right-mover one-particle states.

We study the curvature effect in four dimensions in section \ref{section-fourD}. In section \ref{section-new-action} we introduce a diffeomorphism invariant action with local Lorentz symmetry for the separated spin-up component of free left-handed fermions and derive Eq.\eqref{Int-SF-4D}. Eq.\eqref{Int-switch-ordering} is explained in section \ref{section-Kerr} where we look in on the Kerr geometry. Our results are summarized in  section \ref{section-conclusion}.

\section{General properties of Feynman propagator}\label{section-general-prop}
 Our starting point is Eq.\eqref{1st} in which
    \be
    \cS:=i\int d^Dy\sqrt{-g(y)}\,\overline{\psi}(y)\bD(y)\psi(y),
    \label{Dirac-action-D-dimension}
    \ee
 where $\bD(y)$ denotes the Dirac operator. This equation implies that
    \be
    \bD(x)\psi(x)=-\frac{i}{\sqrt{-g}}\frac{\delta \cS}{\delta\overline{\psi}(x)}.
    \label{Dirac-Eq1}
    \ee
Noting that the action $\cS$ is real-valued, Eq.\eqref{Dirac-action-D-dimension} also gives\footnote{ ${\gamma^a}^\dag=\gamma^0\gamma^a\gamma^0$ and $\gamma^a\gamma^b+\gamma^b\gamma^a=2\eta^{ab}$. Therefore, $\gamma^0\gamma^0=1$ and ${\gamma^0}^\dag=\gamma^0$.}
    \be
    \cS:=-i\int d^Dy\sqrt{-g(y)}\left[\overline{\psi}(y)\gamma^0\overleftarrow{{\bD(y)^\dag}}\right]\gamma^0\psi(y),
    \label{Dirac-action-star-D-dimension}
    \ee
where we have used the complex conjugation rule for Grassmann numbers
    \be
    (c_1c_2)^*=c_2^*c_1^*.
    \ee
Eq.\eqref{Dirac-action-star-D-dimension} gives
    \be
    \overline{\psi}(x)\gamma^0\overleftarrow{{\bD(x)^\dag}}\gamma^0=-\frac{i}{\sqrt{-g}}\frac{\delta \cS}{\delta{\psi}(x)}.
    \label{Dirac-Eq2}
    \ee
Eq.\eqref{Dirac-Eq1} and Eq.\eqref{1st} together give
    \bea
    \bD(x)S_F(x,x')&=&-(-g(x))^{-\frac{1}{2}}\cZ^{-1}\int D\overline\psi D\psi\left( \frac{\delta e^{i\cS}}{\delta\overline{\psi}(x)}\right)\overline\psi(x')\\
    \label{SF-Dx-1}
    &=&(-g(x))^{-\frac{1}{2}}\delta^D(x-x').
    \eea
Eq.\eqref{SF-Dx-1} is obtained by integration by part. Similarly, Eq.\eqref{Dirac-Eq2} and Eq.\eqref{1st}  give
    \be
    \bD(x')\gamma^0S_F(x,x')^\dag\gamma^0=-(-g(x'))^{-\frac{1}{2}}\delta^D(x-x').
    \label{SF-Dx-2}
    \ee
This result together with Eq.\eqref{SF-Dx-1} implies that
    \be
    \bD(x')\left[S_F(x',x)+\gamma^0S_F(x,x')^\dag\gamma^0\right]=0.
    \ee
So,
    \be
     S_F(x',x)+\gamma^0S_F(x,x')^\dag\gamma^0\circeq0,
    \label{identity-general}
    \ee
where  $\circeq$ indicates that a function $\fU(x')$ solving the homogenous field equation $\bD(x')\fU(x')=0$ has been dropped from the right hand side of Eq.\eqref{identity-general}.

 To recognize this result in  the four dimensional Minkowski spacetime we note that Eqs.\eqref{g1} and \eqref{g2} give
    \begin{align}
    &\gamma^0S^A(x,x')^\dag\gamma^0=S^A(x',x),\\
    &\gamma^0S^B(x,x')^\dag\gamma^0=S^B(x',x).
    \end{align}
Consequently, Eq.\eqref{2nd} gives
    \be
     S_F(x',x)+\gamma^0S_F(x,x')^\dag\gamma^0=S^A(x',x)-S^B(x',x).
    \ee

\subsection{Massless Dirac field}
Dirac operator for massless Dirac fields satisfies the identity\footnote{$\gamma^5$ anti-commutes with $\gamma^a$ and ${\gamma^5}\gamma^5=1$. $P_L:=\frac{1-\gamma^5}{2}$ and $P_R:=\frac{1+\gamma^5}{2}$ are projection operator which separate the left-handed and the right-handed components of $\psi=\psi_L+\psi_R$ defined by  $\psi_L:=P_L\psi$ and  $\psi_R:=P_R\psi$.}
    \be
    \gamma^5\bD\gamma^5=-\bD.
    \ee
Thus, $\bD=\bD_L+\bD_R$ where
    \begin{align}
    &\bD_L:=P_R\bD P_L, &\bD_R:=P_L\bD P_R.
    \end{align}
The left-handed and right-handed components of Dirac field decouple consequently, i.e.,
    \be
    \cS=\cS_L+\cS_R,
    \ee
where, for example,
    \be
    \cS_L:= i\int d^Dy\sqrt{-g(y)}\,\overline{\psi_L}(y)\bD_L(y)\psi_L(y).
    \ee
Eq.\eqref{1st} implies that the Feynman propagator decomposes accordingly,
    \be
    S_F(x,x')=S^{L}_F(x,x')+S^{R}_F(x,x'),
    \label{SF-SFL-SFR}
    \ee
where,
    \begin{align}
    \label{SFL-definition}
    &S_F^{L}(x,x'):=P_L S_F(x,x')P_R,\\
     \label{SFR-definition}
    &S_F^{R}(x,x'):=P_R S_F(x,x')P_L.
    \end{align}

In two dimensions, we can choose $\gamma^0={\boldsymbol\sigma}^1$ and $\gamma^5={\boldsymbol\sigma}^3$ where $\boldsymbol{\sigma}^i$ denote the Pauli matrices. Eq.\eqref{SFL-definition} implies that $S_F^{L}(x,x')$  has only one nonzero component which we denote by $S^{+}(x,x')$. Similarly we denote the non-zero component of $S_F^{R}(x,x')$ by $S^{-}(x,x')$. Following Eq.\eqref{SF-SFL-SFR} we obtain
    \be
    S_F(x,x')=\left(\begin{array}{cc}0&S^-_F(x,x')\\S^+_F(x,x')&0\end{array}\right).
    \label{2D-SF-format}
    \ee
Eq.\eqref{identity-general} reads,
    \be
    S_F^{\pm}(x',x)\circeq-S_F^{\pm}(x,x')^*.
     \label{identity-massless-pm}
     \ee

In four dimensions we  choose
    \begin{align}
    &\gamma^0:=\left(\begin{array}{cc}{\bf 0}&{\bf 1}\\{\bf 1} &{\bf 0}\end{array}\right)&\gamma^5:=\left(\begin{array}{cc}{\bf -1}&{\bf 0}\\{\bf 0} &{\bf 1}\end{array}\right),
    \end{align}
Following Eqs.\eqref{SFL-definition} and \eqref{SFR-definition}, we denote the nontrivial components of $S_F^L(x,x')$ and $S_F^R(x,x')$ by $S_F^{(L)}(x,x')$ and $S_F^{(R)}(x,x')$ respectively. Therefore,
    \begin{align}
    &S_F(x,x')=\left(\begin{array}{cc}{\bf 0}&S_F^{(L)}(x,x')\\S_F^{(R)}(x,x') &{\bf 0}\end{array}\right),
    \end{align}
and Eq.\eqref{identity-general} reads, e.g.,
    \be
     S_F^{(L)}(x',x)\circeq-S_F^{(L)}(x,x')^\dag.
    \label{identity-massless}
    \ee
\section{Two dimensional Minkowski spacetime}\label{section-flat}
 In this section we review the second quantization of massless Dirac fermions in inertial reference frames in two dimensional Minkowski spacetime. We show that one-particle states are imprinted, in a retrievable manner, in the  short distance singularity of the Green's function with Feynman boundary condition. We will use this result to justify the definition of left-mover vacuum state and left-mover one-particle states on a time dependent curved background, later in section \ref{section-frame}.

In Minkowski spacetime whose line element is $ds^2=d{x^0}^2-d{x^1}^2$, Dirac field equation is $\gamma^\mu\partial_\mu \psi(x)=0$ where, $\gamma^0:=\boldsymbol{\sigma}^1$ and $\gamma^1:=-i\boldsymbol{\sigma}^2$ are Dirac matrices. Introducing
    \be
    \boldsymbol{\sigma}^\pm:=\frac{1}{2}\left(\boldsymbol{\sigma}^1\pm i\boldsymbol{\sigma}^2\right),
    \ee
Dirac operator can be given as
    \be
     \gamma^\mu\partial_\mu=\sum_{a=\pm}{\boldsymbol\sigma}^{-a}\partial_a=2\left(\begin{array}{cc}0&\partial_-\\\partial_+&0\end{array}\right),
    \label{Dirac-equation-2D-Mink}
    \ee
where
    \begin{align}
    &\partial_a:= \frac{1}{2}\left(\partial_0+a\partial_1\right),&a=\pm,
    \end{align}
correspond to the light-cone coordinates $x^\pm:=x^0\pm x^1$.

We also introduce $\gamma^5:=\gamma^0\gamma^1=\boldsymbol{\sigma}^3$ and use it to decompose the Dirac field $\psi$ into its left-handed and right-handed components, $\psi^+$ and $\psi^-$ respectively,
    \begin{align}
    &\psi^-=\left(\begin{array}{c}\psi_-\\0 \end{array}\right),&\psi^+=\left(\begin{array}{c}0\\\psi_+ \end{array}\right),
    \end{align}
satisfying
     \begin{align}
    &\gamma^5\psi^a=-a\psi^a, &a=\pm.
    \end{align}
 The corresponding field equation is
    \begin{align}
    &\partial_a\psi^{-a}=0,&a=\pm,
    \end{align}
implying that the left-handed component $\psi^+(x)=\psi^+(x^+)$ is left-mover and the right-handed component $\psi^-(x)=\psi^-(x^-)$ is right-mover.

   The quantized massless Dirac field is given by $\psi(x)=\psi^-(x)+\psi^+(x)$ and ${\overline\psi}(x)=\overline{\psi^-}(x)+\overline{\psi^+}(x)$, where
    \begin{align}
    &\psi^a(x):=\int_{-\infty}^\infty \frac{dp}{2\pi}\frac{u^a(p)}{\sqrt{2\left|p\right|}} \left(A_p^ae^{-ip\cdot x}+{B_p^a}^\dag e^{ip\cdot x}\right),\\
    &\overline{\psi^{a}}(x):=\int_{-\infty}^\infty \frac{dp}{2\pi}\frac{{\bar u^{a}}(p)}{\sqrt{2\left|p\right|}}\left({A_p^{a}}^\dag e^{ip\cdot x}+{B_p^{a}} e^{-ip\cdot x}\right),
    \end{align}
  $p\cdot x:=\left|p\right|x^0-p\,x^1$,
    \begin{align}
    &\gamma^5 u^a(p)=-au^a(p),&  a=\pm,
    \label{u-eigen}
    \end{align}
 and $\bar u^a(p):={u^a}(p)^\dag\gamma^0$. $A^a_p$ and ${A^a_p}^\dag$, and $B^a_p$ and ${B^a_p}^\dag$ are the annihilation and creation operators for fermions ($A$-particles) and anti-fermions ($B$-particles) with helicity $a$ and momentum $p$ respectively. The field equation reads
    \begin{align}
    \label{u-field-equation}
    &(\left|p\right|+ap)u^a(p)=0,&a=\pm,
    \end{align}
whose solution is\footnote{We have required that $u^a(p)^\dag u^a(p)=2\left|p\right|\theta(-ap)$. This normalization is reflected in Eq.\eqref{A1} and \eqref{A2}.}
    \begin{align}
    u^a(p)=\sqrt{2\left|p\right|}\theta(-a p)\frac{1-a\gamma^5}{2}\left(\begin{array}{c}1\\1\end{array}\right).
    \label{ua-solution}
    \end{align}
 This result implies that $\psi^-(x)$ annihilates right-moving  fermions and creates right-moving anti-fermions. Similarly, $\psi^+(x)$ annihilates left-moving  fermions and creates left-moving anti-fermions.

The vacuum state $\left|0\right\rangle$ is defined to be the state such that $A_p^a\left|0\right\rangle=B_p^a\left|0\right\rangle=0$.
The one-particle states of fermions and anti-fermions are defined by $\left|A;p,a\right\rangle:=\sqrt{2\left|p\right|}{A_p^a}^\dag\left|0\right\rangle$ and $\left|B;p,a\right\rangle:=\sqrt{2\left|p\right|}{B_p^a}^\dag\left|0\right\rangle$ respectively. These can be used to show that
    \begin{align}\label{A1}
    \left|\overline{\psi^a}(x)\right\rangle=\int_{-\infty}^\infty \frac{dp}{4\pi\left|p\right|}{\bar u^a(p)} e^{ip\cdot x}\left|A;p,a\right\rangle,\\
    \label{A2}
    \left\langle{\psi^a}(x)\right|=\int_{-\infty}^\infty \frac{dp}{4\pi\left|p\right|}{ u^a(p)} e^{-ip\cdot x}\left\langle A;p,a\right|,
    \end{align}
where $\left|\overline{\psi^a}(x)\right\rangle:=\overline{\psi^a}(x)\left|0\right\rangle$ and $\left\langle {\psi^a}(x)\right|:=\left\langle0\right|{\psi^a}(x)$.
Using the orthogonality of the one-particle state, $\left\langle A;p,b|A;p',a\right\rangle=4\pi\left|p\right|\delta(p-p')\delta^{ab}$ one verifies that the matrix element $S^A(x,x'):=\left\langle\psi(x)|\overline{\psi}(x')\right\rangle$ decomposes according to $\bS^A=\bS^{A-}+\bS^{A+}$ whose entries are  given by
    \be
    S^{Aa}_{rs}(x,x')=\int_{-\infty}^\infty \frac{dp}{4\pi\left|p\right|}u^a_r(p){\bar u^a_s(p)} e^{-ip\cdot (x-x')},
   \label{SA}
   \ee
where $u^a_r(p)$ is the $r$-th component of $u^a(p)$. $\bS^{Aa}$ encodes the amplitude for a particle whose helicity is $a$  propagating from $x'$ to $x$. Similarly,
    \begin{align}
    \label{B1}
    &\left|\psi^a(x)\right\rangle=\int_{-\infty}^\infty \frac{dp}{4\pi\left|p\right|}{ u^a(p)} e^{ip\cdot x}\left|B;p,a\right\rangle,\\
    \label{B2}
    &\left\langle \overline{\psi^a}(x)\right|=\int_{-\infty}^\infty \frac{dp}{4\pi\left|p\right|}{ \bar u^a(p)} e^{-ip\cdot x}\left\langle B;p,a\right|,
    \end{align}
where $\left|\psi^a(x)\right\rangle:=\psi^a(x)\left|0\right\rangle$ and $\left\langle \overline{\psi^a}(x)\right|:=\left\langle 0\right|\overline{\psi^a}(x)$.
Once again, the orthogonality relation $\left\langle B;p,b|B;p',a\right\rangle=4\pi\left|p\right|\delta(p-p')\delta^{ab}$ decomposes  the amplitude $S^B(x,x'):=\left\langle\overline{\psi}(x')|\psi(x)\right\rangle$ into its helicity components $\bS^B=\bS^{B-}+\bS^{B+}$ whose entry
    \be
    S^{Ba}_{rs}(x,x')=\int_{-\infty}^\infty \frac{dp}{4\pi\left|p\right|}u^a_r(p){\bar u^a_s(p)} e^{ip\cdot (x-x')},
    \label{SB}
    \ee
gives the amplitude for an antiparticle with helicity $a$ to propagate from $x$ to $x'$.

Since
    \be
    u^a(p){\bar u^a(p)}=2\left|p\right|\theta(-ap){\boldsymbol\sigma}^{-a},
    \label{u-sigma}
    \ee
one verifies that
    \be
    S^{Ba}_{rs}(x',x)=S^{Aa}_{rs}(x,x')=\int_0^\infty\frac{dp}{2\pi}e^{-ip(x^a-{x'}^a)}{\boldsymbol\sigma}^{-a}_{rs}.
    \label{SA-SB}
    \ee
Using the equality
    \be
    \int_0^\infty dp\, e^{-ipx}=\pi\delta(x)-i\int_0^\infty dp\,\sin px,
    \label{delta-identity}
    \ee
in which, $\delta(x):=\frac{d}{dx}\theta(x)$ denotes the Dirac delta function, one obtains
    \be
    S^{Aa}_{rs}(x,x')=\Big\{\frac{1}{2}\delta(x^a-{x'}^a)-\frac{i}{2\pi}\int_0^\infty dp\,\sin\! \left[p(x^a-{x'}^a)\right]\Big\}{\boldsymbol\sigma}^{-a}_{rs}.
    \ee

 The Feynman propagator  is given by
    \be
    {S_{F}}_{rs}:=\theta(x^0-{x'}^0)S^{A}_{rs}(x,x')-\theta({x'}^0-x^0)S^B_{rs}(x,x').
  \label{SF}
    \ee
Therefore, $S_F=S_F^-(x,x'){\boldsymbol\sigma}^{+} +S_F^+(x,x'){\boldsymbol\sigma}^{-}$ where,
    \be
    S_F^ a(x,x')=\frac{1}{2}{\rm sgn}(x^0-{x'}^0)\delta(x^a-{x'}^a)-\frac{i}{2\pi}\int_0^\infty dp\,\sin\! \left[p(x^a-{x'}^a)\right].
    \label{SFa}
    \ee

\subsection{Upside down approach}\label{section-upside-down}
Now we put things the other way around. We compute a Green's function $\cG(x,x')$ for the Dirac operator $\gamma^\mu\partial_\mu$ satisfying  the condition \eqref{identity-massless-pm}, and impose the Feynman boundary condition in order to obtain $S_F(x,x')$. We show that one-particle states and the vacuum state can be recognized in this way.

To this aim, we assume that the Green's function is decomposed into its left-handed and right-handed components $\cG^{(a)}(x,x')$ according to Eq.\eqref{2D-SF-format}, i.e.,
    \be
    \cG(x,x')=\sum_{a=\pm}\cG^{(a)}(x,x'){\boldsymbol\sigma}^{-a}.
    \ee
Consequently,  the field equation \eqref{Dirac-equation-2D-Mink} reads
    \begin{align}
    &2\partial_{-a}\cG^{(a)}(x,x')=\delta^2(x-x'),&a=\pm,
     \end{align}
whose solution is
    \be
    \cG^{(a)}(x,x')\circeq\frac{1}{2}{\rm sgn}(x^0-{x'}^0)\delta(x^a-{x'}^a),
    \label{short}
    \ee
also satisfying  Eq.\eqref{identity-massless-pm}.

The expression \eqref{short} reproduces the first term on the right hand side of Eq.\eqref{SFa}. Recalling that for $x^0>{x'}^0$ and $x^0<{x'}^0$, the Feynman propagator corresponds to the amplitude for particles with positive frequency to propagate from $x'$ to $x$ and from $x$ to $x'$ respectively, the second term on the right hand side of equation \eqref{SFa}, which is a solution to the homogeneous field equation, is uniquely determined. In fact  Eq.\eqref{delta-identity} shows that  this term removes the negative frequencies in the spectrum of $\delta(x^a-{x'}^a)$.

By rewriting the result in terms of the $\theta$ function similarly to Eq.\eqref{SF} we  recognize $\bS^{Aa}$ and $\bS^{Ba}$ as given by Eq.\eqref{SA-SB}. $\boldsymbol\sigma^{a}$ can be factorized identically to  Eq.\eqref{u-sigma} whose solution $u^a(p)$,  given in Eq.\eqref{ua-solution}, is unique up to a phase factor. As a result, the matrix elements $S^{Aa}_{rs}(x,x')$ and $S^{Ba}_{rs}(x,x')$ can be recognized as given by Eq.\eqref{SA} and Eq.\eqref{SB} respectively. Finally, they can be factorized according to  Eqs.\eqref{A1} and \eqref{A2} and Eqs.\eqref{B1} and \eqref{B2} respectively. At this point we postulate the orthogonal one-particle states $\left|A;p,a\right\rangle$ and $\left|B;p,a\right\rangle$ and subsequently postulate the vacuum state $\left|0\right\rangle$.

\section{Two dimensional curved spacetime}\label{section-fermion}

 We denote the spacetime  metric by $g_{\mu\nu}$  and the Minkowski metric by\footnote{We are describing the Minkowski spacetime in the light-cone gauge. We also recall that the classical trajectory of massless particles are light-like.}
    \be
    \eta=\boldsymbol{\sigma}_1.
   \label{eta-2d=}
    \ee
 The local frame is identified by the tetrad ${e_\mu}^{(a)}$ and its inverse
    \be
    {e^\mu}_{(a)}=\sum_{b=\pm}\eta_{ab}\,g^{\mu\nu}{e_\nu}^{(b)},
     \ee
 satisfying the identity
    \be
    g^{\mu\nu}={e^\mu}_{(+)}{{e^\nu}}_{(-)}+{e^\mu}_{(-)}{{e^\nu}}_{(+)},
     \ee
 where, $g^{\mu\nu}$ is the inverse of $g_{\mu\nu}$. We use the Einstein summation notation when we sum over spacetime indices  $\mu,\nu=0,1$. In the flat spacetime limit, where
    \be
    ds^2=d{x^0}^2-d{x^1}^2,
    \ee
 we choose the inertial frame
     \begin{align}
     &{e^0}_{(a)}=\frac{1}{\sqrt 2}, &{e^1}_{(a)}=-a\frac{1}{\sqrt 2},
     \label{standard-inertial-frame-2D}
     \end{align}
 where ${e^\mu}_{(\pm)}:=\frac{1}{\sqrt 2}({e^\mu}_{(0)}\pm{e^\mu}_{(1)})$. We assume that $g_{\mu\nu}$ is a continuous function of the spacetime coordinates,  $g:=\det g_{\mu\nu}<0$ and $g_{11}<0$. That is to say, the spacetime is orientable and   $dx^0=0$ corresponds to space-like intervals. Thus $g^{00}>0$ and we choose the tetrad such that ${e^0}_{(a)}>0$. Furthermore, we select the $\pm$ sign such that
    \be
    {e_0}^{(+)}{e_1}^{(-)}>{e_0}^{(-)}{e_1}^{(+)},
    \ee
 and consequently
    \be
    \sqrt{- g}={e_0}^{(+)}{e_1}^{(-)}-{e_0}^{(-)}{e_1}^{(+)}.
    \ee
 As a result
    \begin{align}
    \label{E-e}
    &{E^0}_{(a)}=a {e_1}^{(-a)},     &{E^1}_{(a)}=-a {e_0}^{(-a)},
    \end{align}
 where ${E^\mu}_{(a)}:=\sqrt{- g}\,{ e^\mu}_{(a)}$.

 The action  is given by \cite{Alvaerz-Gaume-Witten}
     \be
     \cS=\frac{i}{\sqrt{2}}\sum_{a=\pm}\int d^2x\, {E^\mu}_{(a)}(x)\left(\overline\psi(x)\gamma^a\overleftrightarrow{\p_\mu}\psi(x)\right),
     \label{Dirac-action}
     \ee
where,
    \be
    \gamma^a:=\frac{1}{2}(\gamma^0-a\gamma^1)={\boldsymbol\sigma}^a,
     \ee
and the operator $\overleftrightarrow\partial$ is defined according to the rule
    \be
    \varphi_1(x)\overleftrightarrow{\p_\mu} \varphi_2(x):=-(\partial_\mu\varphi_1(x))\varphi_2(x)+\varphi_1\partial_\mu\varphi_2(x).
    \ee
Assuming that
     \begin{align}
     \label{psi-components}
     & \psi=\left(\begin{array}{c}\psi_-\\ \psi_+\end{array}\right),
     \end{align}
where $\psi_+$ and $\psi_-$ are the so-called left-handed spinor and the right-handed spinor respectively, Eq.\eqref{Dirac-action} reads
    \be
    \cS=\cS_++\cS_-,
    \ee
where
    \bea
    \label{Dirac-action-a-1st}
     \cS_a&:=&\frac{i}{\sqrt{2}}\int d^2x\, {E^\mu}_{(a)}(x)\left(\psi_a(x)^*\overleftrightarrow{\p_\mu}\psi_a(x)\right)\\
     \label{Dirac-action-a}
     &=&i\int d^2x\,\sqrt{-g(x)} \psi_a(x)^*\cD_{-a}\psi_a(x),
     \eea
in which, we have integrated by part and dropped a boundary term to obtain the second equality, and
    \be
    \cD_{-a}(x):= \sqrt{\frac{-2}{g(x)}}\left({E^\mu}_{(a)}(x)\p_\mu+\frac{1}{2}\p_\mu {E^\mu}_{(a)}(x)\right),
    \label{D-a}
    \ee
 is the Dirac operator in the corresponding sector. In the flat spacetime limit, and in the inertial frame \eqref{standard-inertial-frame-2D}
    \begin{align}
    &\cD_{a}=2\partial_a,&a=\pm.
     \end{align}

  This theory is invariant under local `Lorentz' transformations\footnote{Recall that we are using the light-cone coordinates. The fastest way to recognize the generator of Lorentz transformation in this representation is to note that for spinors, the generator is given by $\frac{i}{4}[\gamma^0,\gamma^1]=\frac{i}{2}{\boldsymbol\sigma}_3$. Thus for spacetime coordinates, i.e., in the real spin 1 representation, it is given by ${\boldsymbol\sigma}_3$. }
     \begin{align}
     \label{V-transforation-E}
     & {e_\mu}^{(a)}(x)\to {^{(\lambda)}\!e_\mu}^{(a)}(x)=e^{a\lambda(x)}{e_\mu}^{(a)}(x),\\
     \label{V-transforation-theta}
     &\psi_a(x)\to {^{(\lambda)}\!\psi}_a(x)=e^{a{\frac{\lambda(x)}{2}}}\psi_a(x).
     \end{align}
The solution to the classical field equation $\cD_{-a}(x)\psi_{a}(x)=0$ is given by
    \be
    \psi_a(x):={\fu}^{(a)}\left(z^{(a)}(x)\right)e^{-\cR^{(a)}(x)},
    \label{classical-solution}
    \ee
 in which $\fu^{(a)}$ is a smooth function, and  $z^{(a)}$ and $\cR^{(a)}$ are real-valued functions. $\cR^{(a)}$ solves
    \begin{align}
    \label{R=}
    &{E^\mu}_{(a)}(x)\partial_\mu \cR^{(a)}(x)=\frac{1}{2}\p_\mu {E^\mu}_{(a)}(x),
    \end{align}
which according to the Peano existence theorem has at least one solution only if ${E^\mu}_{(a)}(x)$ are continuous \cite{Peano}. $z^{(a)}$ are given by the following equations
    \begin{align}
    \label{z0=}
    &\partial_0 z^{(a)}=-a{E^1}_{(a)}e^{-2\cR^{(a)}},\\
    \label{z1=}
    &\partial_1 z^{(a)}=a{E^0}_{(a)}e^{-2\cR^{(a)}}.
    \end{align}
Eq.\eqref{R=} implies that $[\partial_0,\partial_1]z^{(a)}=0$, which is the necessary condition for the existence of $z^{(a)}$, and
    \begin{align}
    \label{z=}
    &{E^\mu}_{(a)}(x)\partial_\mu z^{(a)}(x)=0.
    \end{align}
Using Eq.\eqref{E-e}, one can rewrite Eqs.\eqref{z0=} and \eqref{z1=} in the following way,
    \begin{align}
    \label{z0z1=e}
    &\partial_\mu z^{(a)}={e_\mu}^{(-a)}e^{-2\cR^{(a)}},&\mu=0,1,
    \end{align}
which together with Eq.\eqref{eta-2d=} gives the celebrated result that two dimensional manifolds are conformally flat,
    \be
    ds^2=2 e^{2\cR^{(+)}}e^{2\cR^{(-)}}dz^{(+)}dz^{(-)}.
    \ee

  \subsection{Feynman propagator}\label{section-Green}
  In section \ref{section-general-prop} we observed that the Feynman propagator $S_F^a(x,x')$ solves the equation
    \be
    \cD_{-a}(x)S_F^a(x,x')=\frac{1}{\sqrt{-g(x)}}\delta^2(x-x'),\  \ \ a=\pm,
   \label{start}
    \ee
  and simultaneously Eq.\eqref{identity-massless-pm}. Define
    \be
    \delta^{(a)}_T(x,x'):=2^{-\frac{3}{2}}{\rm sgn}(x^0-{x'}^0)\delta\!\left(z^{(a)}(x)-z^{(a)}(x')\right).
    \ee
 Eq.\eqref{z=} and Eq.\eqref{R=}  give
    \begin{align}
    &\cD_{-a}(x)\left(\delta^{(a)}_T(x,x')e^{-\cR^{(a)}(x)}\right)=\nn\\
    &({-g(x)})^{-\frac{1}{2}}{E^0}_{(a)}(x')\delta(x^0-{x'}^0)\delta\!\left(z^{(a)}(x)-z^{(a)}(x')\right)e^{-\cR^{(a)}(x)},
    \label{App-Green-1}
    \end{align}
 and Eq.\eqref{z1=} gives,
    \be
    \delta(x^0-{x'}^0)\,\delta\!\left(z^{(a)}(x)-z^{(a)}(x')\right)=\delta(x^0-{x'}^0)\frac{\delta(x^1-{x'}^1)}{{E^0}_{(a)}(x')}e^{2\cR^{(a)}(x')},
    \label{result}
    \ee
 where we have noticed that  ${E^0}_{(a)}>0$. Therefore,
    \be
   S_F^a\left(x,{x'}\right)\circeq\delta^{(a)}_T(x,x')e^{-\cR^{(a)}(x)-\cR^{(a)}(x')},
    \label{Green-raw}
    \ee

 Since $z^{(a)}(x,x')=0$ indicates a light-like curve, $\delta^{(a)}_T(x,x')$ in Eq.\eqref{Green-raw} can be interpreted in terms of the amplitude corresponding to the propagation of light-like modes similarly to the flat spacetime discussed in section \ref{section-upside-down}.  In this way, the factor $\exp(-\cR^{(a)}(x))$ should be considered as an $x$-dependent normalization, which obstructs an interpretation of the amplitude in terms of one-particle states. Solutions to the homogeneous field equation do not cancel out nonzero $\cR^{(a)}(x)$. In the next subsection we show that this effect is  a reflection of the spacetime curvature.

\subsection{Curvature}\label{section-curvature}
In this subsection, we study the spin connection in $D$-dimensional spacetimes.  For simplicity, here and also later in sections \ref{section-fourD} and \ref{section-new-action}, we  use the Einstein summation notation  when we sum over frame indices $a,b,c,d=0,\cdots,D$.

Dirac operator is given by
    \be
    \bD(e):={e^\mu}_{(a)}\gamma^a\partial_\mu+\boldsymbol{\Omega},
    \label{Dirac-operator}
    \ee
where
    \be
    \boldsymbol{\Omega}:=-\frac{i}{2}{e^\mu}_{(a)}\gamma^a\Omega_\mu(e),
    \label{spin-connection}
    \ee
denotes the spin connection. Here,
    \be
    \Omega_\mu(e):={e^\nu}_{(a)}\nabla_\mu {e_\nu}_{(b)}\Sigma^{ab},
    \label{Omega-mu}
    \ee
$\nabla_\mu$ denotes the Levi-Civita connection, and $\Sigma^{ab}:=\frac{i}{4}[\gamma^a,\gamma^b]$ satisfying the Lorentz algebra
    \be
    i[\Sigma^{ab},\Sigma^{cd}]=\left(\eta^{ac}\Sigma^{bd}+\eta^{bd}\Sigma^{ac}\right)-(a\leftrightarrow b).
    \ee
Under a local Lorentz transformation given by
    \be
    U(\xi):=\exp\left(-\frac{i}{2}\xi_{ab}\Sigma^{ab}\right),
    \label{U-xi}
    \ee
we have $U(\xi)^{-1}\gamma^aU(\xi)={\Lambda(\xi)^a}_b\gamma^b$, where $\Lambda(\xi):=\exp\left(-\frac{i}{2}\xi_{ab}J^{ab}\right)$.    $J^{ab}$, whose entries are given by ${[J^{ab}]^c}_d=i(\eta^{ac}\delta^b_d-\eta^{bc}\delta^a_d)$, satisfy the Lorentz algebra similarly to $\Sigma^{ab}$. One can show that
    \be
    \Omega_\mu(^{(\xi)}\!e)=U(\xi)\Omega_\mu U(\xi)^{-1}+\Xi(\xi)_\mu,
    \label{4d-frame-change}
    \ee
where $\Xi(\xi)_\mu:=2iU(\xi)\partial_\mu U(\xi)^{-1}$ and
    \be
    ^{(\xi)}\!{e^\mu}_{(a)}:={\Lambda(\xi)_a}^b{e^\mu}_{(b)}.
    \label{Lorentz-e}
     \ee
Consequently,
    \be
    \bD(^{(\xi)}\!e)=U(\xi)\bD(e)U(\xi)^{-1}.
    \label{Lorentz-D}
    \ee
 Let
    \be
    \mho_{\mu\nu}(\Omega):=\nabla_\mu\Omega_\nu-\nabla_\nu\Omega_\mu-\frac{i}{2}[\Omega_\mu,\Omega_\nu].
    \ee
Since $\mho_{\mu\nu}\left(\Xi(\xi)\right)=0$, and $\mho_{\mu\nu}(\Omega(e))=R_{\mu\nu\rho\sigma}{e^\rho}_{(a)}{e^\sigma}_{(b)}\Sigma^{ab}$, in which $R_{\mu\nu\rho\sigma}$ denotes the Riemann tensor, we conclude that in curved spacetime, $\boldsymbol{\Omega}$ can not be eliminated by  local Lorentz transformations.

In two dimensions,
    \be
    R_{\mu\nu\rho\sigma}=\frac{R}{2}(g_{\mu\rho}g_{\nu\sigma}-g_{\mu\sigma}g_{\nu\rho}),
    \ee
where $R$ is the Ricci scalar. Consequently, $\mho_{01}(\Omega(e))=iR\sqrt{-g}\boldsymbol{\sigma}_3$. Furthermore, $\boldsymbol{\Omega}$ is a 2 by 2 anti-diagonal matrix whose non-zero entries are
    \begin{align}
    &\boldsymbol{\Omega}_{12}=\nabla_\mu {e^\mu}_{(-)},\\
    &\boldsymbol{\Omega}_{21}=\nabla_\mu {e^\mu}_{(+)}.
     \end{align}
Henceforth, we denote  it by
    \be
    \boldsymbol{\Omega}=[[\nabla_\mu {e^\mu}_{(-)},\nabla_\mu {e^\mu}_{(+)}]]_{/}.
    \ee
Noting that
    \be
    \partial_\mu {E^\mu}_{(a)}=\sqrt{-g}\nabla_\mu{e^\mu}_{(a)},
    \label{cov-diff}
    \ee
one verifies that $\bD(e)=[[\cD_+,\cD_-]]_/$.

\subsection{Left-handed frames}\label{section-frame}
Eq.\eqref{cov-diff} implies that  under a general coordinate transformation,
    \be
    \partial_\mu {E^\mu}_a(x)\to\partial'_\mu {{E'}^\mu}_{(a)}(x')=\left|\det\!\left(\frac{\partial x}{\partial x'}\right)\right|\partial_\mu {E^\mu}_{(a)}(x),
    \label{g.c.t-fra}
    \ee
in which the determinant is the Jacobian  of the coordinate transformation. Eq.\eqref{V-transforation-E} gives the  local Lorentz transformation rule, $\partial_\mu {E^\mu}_{(a)}(x)\to\partial_\mu {^{(\lambda)}\!E^\mu}_{(a)}(x)$ where,
    \begin{align}
    &e^{a\lambda(x)}\partial_\mu {^{(\lambda)}\!E^\mu}_{(a)}(x)=
    \partial_\mu {E^\mu}_{(a)}(x)-a{E^\mu}_{(a)}(x)\partial_\mu\lambda(x).
    \label{Lorentz}
    \end{align}
The local Lorentz transformation can be used to satisfy  the {\em left-mover gauge} $\partial_\mu{E^\mu}_{(+)}(x)=0$. The transformation rule \eqref{g.c.t-fra} implies that such  {\em left-mover frames} are independent of the coordinate system.

Explicitly, if we give the metric in the conformal gauge,
    \be
    ds^2=e^{2\omega}dx^+dx^-,
    \ee
the left-handed frame is described by
    \begin{align}
    \label{LHF+}
    &{e^-}_{(+)}=\sqrt 2e^{-2\omega},&{e^+}_{(+)}=0,\\
    &{e^-}_{(-)}=0,&{e^+}_{(-)}=\sqrt 2,
    \end{align}
 $\cR^{(+)}(x)=0$ and  $\cR^{(-)}(x)=\omega(x)$.

Using \eqref{LHF+} in Eq.\eqref{Dirac-action-a-1st} gives
    \be
     \cS_+=i\int d^2x\, \psi_+(x)^*\overleftrightarrow{\p_-}\psi_+(x),
     \label{Dirac-action-+}
     \ee
 and consequently, Eq.\eqref{1st} implies that the Feynamnn propagator is given by
    \be
    S_F^ +(x,x')=\frac{1}{2}{\rm sgn}(x^0-{x'}^0)\delta(x^+-{x'}^+)-\frac{i}{2\pi}\int_0^\infty dp\,\sin\! \left[p(x^+-{x'}^+)\right],
    \label{SF+}
    \ee
  similarly to  Eq.\eqref{SFa}. Equivalently, using \eqref{LHF+} in Eq.\eqref{D-a}  we obtain
    \be
    \cD_-=2e^{-2\omega}\partial_-=e^{-2\omega}\left(\partial_0-\partial_1\right),
    \ee
  and Eq.\eqref{SF+} solves Eq.\eqref{start} and Eq.\eqref{identity-massless-pm}. Following section \ref{section-upside-down}, we can interpret $S_F^ +(x,x')$  as the propagation amplitude of fermionic one-particle states.

  Furthermore, following the standard approach to the conformal field theory in two dimensions, one can identify the left-movers in the left-mover gauge as a chiral $c=\frac{1}{2}$ conformal field theory \cite{Ginsparg,Ketov}. Eq.\eqref{Lorentz} implies that the chiral symmetry  is generated by $\lambda^{(+)}$ satisfying the condition ${E^\mu}_{(+)}\partial_\mu\lambda^{(+)}=0$, i.e., $\lambda^{(+)}=\lambda^{(+)}(x^+)$\ \cite{Loran2016}. This theory is the typical example of systems with gravitational anomaly \cite{Alvaerz-Gaume-Witten,Bardeen-Zumino,Alvarez-Gaume-Ginsparg,Leutwyler85}.

\section{Four dimensional curved spacetime}\label{section-fourD}
  Following \cite{Peskin-book} we denote the Minkowski metric by $\eta={\rm diag}(1,-1,-1,-1)$, and use Dirac gamma matrices $\gamma^a:=[[\boldsymbol{\sigma^a},{\boldsymbol{\bar\sigma}^a}]]_{/}$ and $\gamma^5:=[[-{\bf 1}, {\bf 1}]]_\backslash$. The notation we are using in writing $\gamma^5$ indicates that it is block-diagonal. In writing $\gamma^a$ we have considered them as 2 by 2 anti-diagonal matrices whose entries are the 2 by 2 matrices  $\boldsymbol{\sigma^0}:=\boldsymbol{\bar\sigma}^0:={\bf1}$ and ${\boldsymbol{\bar\sigma}}^i:=-\boldsymbol{\sigma}^i$. In this notation,  Eq.\eqref{Omega-mu} reads
    \be
    \Omega_\mu=[[\Omega_{L\mu},\Omega_{R\mu}]]_\backslash,
     \label{Omega-mu-LR}
     \ee
in which  $\Omega_{R\mu}:=-\boldsymbol{\sigma}^2\Omega_{L\mu}^*\boldsymbol{\sigma}^2$. Furthermore, Eq.\eqref{U-xi} reads
    \be
    U(\xi)=[[U_L(\xi),U_R(\xi)]]_\backslash,
    \label{U-LR}
    \ee
 where
    \be
    U_R(\xi)=\boldsymbol{\sigma}^2U_L(\xi)^*\boldsymbol{\sigma}^2.
    \label{UR-UL*}
    \ee
 In  section \ref{section-curvature} we showed that in a curved spacetime,
    \be
    \Omega_\mu\neq 2iU(\xi)^{-1}\partial_\mu U(\xi),
    \ee
 so, Eqs.\eqref{Omega-mu-LR} and \eqref{U-LR} imply that in four dimensional curved backgrounds, there is no local Lorentz transformation $\Lambda(\xi)$ such that $\Omega_{L\mu}(^\xi\!e)=0$. In other words,  the curvature effect can not be removed from the left-handed  sector by means of  local Lorentz transformations.

  Eq.\eqref{spin-connection} gives $      \boldsymbol\Omega=[[\boldsymbol\Omega_R,\boldsymbol\Omega_L]]_/$,
  where $\boldsymbol\Omega_R:={\boldsymbol\sigma}^2\boldsymbol\Omega_L^*{\boldsymbol\sigma}^2$,
      \begin{align}
      \label{Omega-L}
      &\boldsymbol\Omega_L:=\zeta_{(a)}\boldsymbol{\bar\sigma}^a,\\
      \label{zeta-a}
      &\zeta_{(a)}:=\frac{1}{2}\nabla_\mu{e^\mu}_{(a)}+\frac{i}{4}{e^\mu}_{(a)}I_\mu,\\
        \label{I-mu}
      &I_\mu:={\epsilon_a}^{bcd}{e_\mu}^{(a)}{e^\rho}_{(b)}{e^\nu}_{(c)}\partial_\rho{e_\nu}_{(d)},
      \end{align}
  and ${\epsilon_a}^{bcd}:=\eta_{ae}\epsilon^{ebcd}$ in which $\epsilon^{abcd}$ is the totally antisymmetric tensor so that ${\epsilon}^{0123}=1$.  $I_\mu(x)$ is frame dependent. That is, for a local Lorentz transformation $\Lambda(\xi)$, Eq.\eqref{Lorentz-e} gives
    \be
    ^\xi\! I_\mu=I_\mu+{\epsilon_a}^{bcd}\eta_{cn}{e_\mu}^{(a)}{e^\rho}_{(b)}{\Lambda^p}_d\partial_\rho{\Lambda_p}^n,
    \ee
   so we can choose a frame in which $I_\mu=0$. In  the  normal neighborhood of $x'=0$ \cite{Bunch1979},
    \be
    g_{\mu\nu}(x)=\eta_{\mu\nu}-\frac{1}{3}R_{\mu\alpha\nu\beta}x^\alpha x^\beta-\frac{1}{6}\nabla_\gamma R_{\mu\alpha\nu\beta}x^\alpha x^\beta x^\gamma+\cO(x^4),
    \ee
this frame is given by
     \be
     {e_\mu}^a(x)=\delta_\mu^a-\frac{1}{6}{R^a}_{\alpha\mu\beta} x^\alpha x^\beta-\frac{1}{12}\nabla_\gamma {R^a}_{\alpha\mu\beta}x^\alpha x^\beta x^\gamma+\cO(x^4).
    \ee
as can be verified by using the Bianchi identities for the Riemann tensor to show that $I_\mu(x)=\cO(x^3)$.

  Eq.\eqref{Dirac-operator} reads  $\bD(e)=[[\bD_R,\bD_L]]_/$ in which $\bD_R:=\boldsymbol{\sigma}^2\bD_L^*\boldsymbol{\sigma}^2$ and
    \be
    \bD_L:=\boldsymbol{\bar\sigma}^a{e^\mu}_{(a)}\partial_\mu+\boldsymbol{\Omega}_L=\boldsymbol{\bar\sigma}^a\left({e^\mu}_{(a)}\partial_\mu+\zeta_{(a)}\right),
     \label{DL}
     \ee
where we have used Eq.\eqref{Omega-L}. The Dirac field equation $\bD(e)\Psi=0$ gives
    \begin{align}
    \label{DL-psiL-eom}
      &\bD_L\psi_L=0,\\
    &\bD_R\psi_R=0,
    \end{align}
  where $\psi_L$ and $\psi_R$ are the left-handed and right-handed components of $\Psi$ separated by the projection operators $P_L=\frac{ 1-\gamma^5}{2}$ and $P_R=\frac{ 1+\gamma^5}{2}$.

    \subsection{Conformally flat spacetimes}
Suppose
    \be
    ds^2=e^{2\omega(x)}\eta_{\mu\nu}dx^\mu dx^\nu,
    \ee
and choose the tetrad,
    \be
    {e_\mu} ^{(a)}(x)=e^{\omega(x)}{\delta_\mu}^a.
    \label{conf-1st-frame}
    \ee
Eq.\eqref{zeta-a} gives
    \be
    \zeta_{(a)}=\frac{3}{2}e^{-\omega(x)}{\delta^\mu}_a\partial_\mu\omega(x),
    \ee
and Eq.\eqref{DL} gives
    \be
    \bD_L(x)=\boldsymbol{\bar\sigma}^a{\delta^\mu}_a e^{-\omega(x)}\left(\partial_\mu+\frac{3}{2}\partial_\mu\omega(x)\right).
    \label{conf-DL-1}
    \ee
Consider the two-point function
    \be
   \cG_{\rm cf}^{(L)}(x,x'):=i\int \frac{d^4p}{(2\pi)^4}\frac{p\cdot\boldsymbol\sigma }{p\cdot p+i\epsilon}e^{-ip\cdot \left(x-x'\right)}e^{-\frac{3}{2}\left(\omega(x)+\omega(x')\right)},
    \label{conf-flat-green}
    \ee
where
    \be
    p\cdot q:=p_a q^a,
    \ee
and $x^a:={\delta_\mu}^a\, x^\mu$. Using the identity
    \be
    {\boldsymbol\sigma}\cdot p\, {\boldsymbol{\bar\sigma}}\cdot p=p\cdot p,
    \ee
we find that
    \be
    \bD_L(x) \cG_{\rm cf}^{(L)}(x,x')=\frac{\delta^4(x-x')}{\sqrt{-g(x)}}.
    \label{conf-flat-1}
    \ee
Furthermore,
    \be
     \cG_{\rm cf}^{(L)}(x',x)\circeq-\cG_{\rm cf}^{(L)}(x,x')^\dag.
    \label{conf-flat-2}
    \ee

Following the argument in section \ref{section-general-prop}, Eqs.\eqref{conf-flat-1} and \eqref{conf-flat-2} show that
    \be
    S_F^{(L)}(x,x')\circeq \cG_{\rm cf}^{(L)}(x,x').
    \ee
This result can be interpreted as the propagation amplitude of one-particle states with an $x$-dependent normalization because
    \be
    \cG_{\rm cf}^{(L)}(x,x')=\cG^{(L)}_{1\rm p}(x,x') e^{-\frac{3}{2}\left(\omega(x)+\omega(x')\right)},
    \label{conf-SF-standard}
    \ee
where
    \be
    \cG^{(L)}_{1\rm p}(x,x'):= i\int \frac{d^4p}{(2\pi)^4}\frac{p\cdot\boldsymbol\sigma }{p\cdot p+i\epsilon}e^{-ip\cdot \left(x-x'\right)},
    \ee
gives the Feynman propagator in an inertial frame in Minkowski spacetime.\footnote{$\cG^{(L)}_{1\rm p}(x,x')$ is given by the right hand side of Eq.\eqref{2nd} after setting $m=0$ in Eqs.\eqref{g1} and \eqref{g2} and replacing $\psl$ by $p\cdot\boldsymbol\sigma$ therein.}
Such an interpretation is supported by the classical symmetry of massless Dirac fields in four dimensions under the conformal map
    \be
    (e^{2\omega}\eta_{\mu\nu},\psi)\to(\eta_{\mu\nu}, e^{\frac{3}{2}\omega}\psi).
    \ee

The left handed-field $\psi_L$ can be further decomposed into its {\em spin-up} and {\em spin-down} components by means of the projection operators
    \begin{align}
    &\hat p_\uparrow\equiv\frac{1+{\boldsymbol\sigma}_3}{2},&\hat p_\downarrow\equiv\frac{1-{\boldsymbol\sigma}_3}{2},
    \end{align}
which  separate the spin-up component and the spin-down component of the field along the third direction respectively. That is to say, $\psi_L=\psi_L^\uparrow+\psi_L^\downarrow$, where
    \be
    \psi_L^\uparrow:=\hat p_\uparrow\psi_L,
    \label{psiLup-definition}
    \ee
and $\psi_L^\downarrow:=\hat p_\downarrow\psi_L$.

Assuming that $\psi_L^\downarrow=0$, the field equation \eqref{DL-psiL-eom} gives $\bD_L\psi_L^\uparrow=0$ and consequently
    \begin{align}
    \label{conf-spin-up-eom-1}
     &\left(\hat p_\uparrow\bD_L\hat p_\uparrow\right)\psi_L^\uparrow=0.
    \end{align}
 Eq.\eqref{conf-spin-up-eom-1} can be considered as the classical field equation corresponding to the action,
     \be
     \cS[{\psi_L^\uparrow}^\dag,\psi_L^\uparrow]=i\int d^4 x\sqrt{-g(x)}{\psi_L^\uparrow}^\dag \bD_L\psi_L^\uparrow.
      \label{conf-action}
      \ee
 The corresponding Feynman propagator $\bS_F^{\uparrow(L)}(x,x')$ can be computed by the path integral \eqref{1st}. Eq.\eqref{1st} and Eq.\eqref{psiLup-definition}  already imply that $\bS_F^{\uparrow(L)}(x,x')$ has only one nonzero component which we denote by  $S_F^{\uparrow(L)}(x,x')$. More explicitly,
    \be
    \bS_F^{\uparrow(L)}(x,x')= {\hat p_\uparrow}\bS_F^{\uparrow(L)}(x,x'){\hat p_\uparrow}=  S_F^{\uparrow(L)}(x,x'){\hat p_\uparrow}.
    \ee
Similarly we use the symbol $D_L^\uparrow$ to denote the nonzero component of $\hat p_\uparrow\bD_L\hat p_\uparrow$.
Following section \ref{section-general-prop} we know that the Feynman propagator $S_F^{\uparrow(L)}(x,x')$ satisfies the following equations
    \begin{align}
    \label{conf-SF-cond-1}
    &D_L^\uparrow S_F^{\uparrow(L)}(x,x')=(-g(x))^{-\frac{1}{2}}\delta^4(x-x'),\\
    \label{conf-SF-cond-2}
    &S_F^{\uparrow(L)}(x',x)\circeq-S_F^{\uparrow(L)}(x,x')^*.
    \end{align}
 Eq.\eqref{conf-DL-1} implies that  in the frame of reference \eqref{conf-1st-frame}
    \be
    D_L^\uparrow=2e^{-\omega(x)}\left(\partial_-+\frac{3}{2}\partial_-\omega(x)\right),
    \ee
where $x^\pm:=x^0\pm x^3$. Consequently
    \be
    S_F^{\uparrow(L)}(x,x')\circeq S_F^+(x,x')\delta^2( {\bf x}_\bot- {{\bf x}_\bot'})e^{-\frac{3}{2}(\omega(x)+\omega(x'))},
   \label{conf-SF-nonstandard}
   \ee
in which
    \begin{align}
    \delta^2({\bf x}_\bot-{\bf x}_\bot'):=\delta(x^1-{x'}^1)\delta(x^2-{x'}^2),
    \end{align}
and  $S_F^+(x,x')$ is given in Eq.\eqref{SFa}.  Eq.\eqref{conf-SF-nonstandard} implies that $S_F^{\uparrow(L)}(x,x')$ corresponds to propagating modes confined to a two dimensional subspace. Since  $S_F^+(x,x')$ is the Feynman propagator of left-moving spinors in a two dimensional Minkowski spacetime we conclude that  similarly to the standard result \eqref{conf-SF-standard}, the Feynman propagator \eqref{conf-SF-nonstandard} can be interpreted in terms of one particle states with an $x$-dependent normalization. So we sacrificed the spin-down component of the field but gained nothing new in the frame of reference \eqref{conf-1st-frame}.

Now consider another frame of reference given by
    \begin{align}
    \label{conf-flat-LHF-1}
     &{e_\mu}^{(+)}dx^\mu=\frac{1}{2}e^{-2\omega}dx^+,\\
     \label{conf-flat-LHF-2}
     &{e_\mu}^{(-)}dx^\mu=e^{4\omega}dx^-,\\
     \label{conf-flat-LHF-3}
     &{e_\mu}^{(a)}dx^\mu=e^\omega dx^a,&a=1,2,
    \end{align}
This implies that the Minkowski metric in the local frame is given in the light-cone gauge
    \begin{align}
    \eta_{-+}=1,   & &\eta_{\pm a}=0,  &  &\eta_{ab}=-\delta_{ab},
    \label{light-cone-4D}
    \end{align}
for $a,b=1,2$. Eq.\eqref{zeta-a} gives $\zeta_{(-)}=0$,\footnote{\label{footnote-Imzeta}Both of Eqs.\eqref{conf-flat-LHF-1} and  \eqref{conf-flat-LHF-2} give ${\sqrt{-g}}{e^-}_{(-)}=1$ and ${e^\mu}_{(-)}=0$ for $\mu=+,1,2$. Thus, ${\rm Re}\,\zeta_{(-)}$ which is proportional to $\partial_\mu(\sqrt{-g}{e^\mu}_{(-)})$ is zero. The imaginary part of $\zeta_{(-)}$ is given by ${\epsilon_-}^{bcd}{e^\mu}_{(b)}{e^\nu}_{(c)}\nabla_\mu e_{\nu(d)}$. Eq.\eqref{conf-flat-LHF-3} implies that the contribution from the $b=-$ terms in ${\rm Im}\,\zeta_{(-)}$ is zero. Now consider the contribution from the $c=-$ terms. Eqs.\eqref{conf-flat-LHF-1} and \eqref{conf-flat-LHF-2} imply that only $d=+$ contributes in ${e^\nu}_{(-)}\nabla_\mu e_{\nu(d)}$. Therefore, the $c=-$ terms (and similarly the $d=-$ terms) add zero to ${\rm Im}\,\zeta_{(-)}$.} thus
    \be
    D_L^\uparrow= e^{-4\omega}\left(\partial_0-\partial_3\right),
    \label{conf-flat-Dirac-operator-LHF}
    \ee
and Eqs.\eqref{conf-SF-cond-1} and \eqref{conf-SF-cond-2} give
    \be
    S_F^{\uparrow(L)}(x,x')=S_F^+(x,x')\delta^2({\bf x}_\bot-{\bf x}_\bot').
    \label{conf-SF}
    \ee
This can be also verified by using the path integral \eqref{1st} and noting that in this frame of reference, Eq.\eqref{conf-action} reads
    \be
     \cS[{\psi_L^\uparrow}^\dag,\psi_L^\uparrow]=2i\int d^4 x{\psi_L^\uparrow}^\dag \partial_-\psi_L^\uparrow.
      \label{conf-action-1}
      \ee
Following the argument in section \ref{section-upside-down} $S_F^{\uparrow(L)}(x,x')$ in Eq.\eqref{conf-SF} can be interpreted as the propagation amplitude of one-particle states localized on a two dimensional Minkowski spacetime.

In summary, in the model \eqref{conf-action} for the  separated  spin-up component of the left-handed Dirac field,  the Feynman propagator can be described in terms of propagating modes localized on a two dimensional subspace and there exists a frame of reference in which the one-particle states can be defined similarly to the second quantization in  two dimensional Minkowski spacetime. In fact, in this model, the frame given by Eqs.\eqref{conf-flat-LHF-1}, \eqref{conf-flat-LHF-2} and \eqref{conf-flat-LHF-3} is reminiscent of the left-handed frame in two dimensions.

 In the next section we show that the action \eqref{conf-action}  enjoys local Lorentz symmetry in addition to the diffeomorphism invariance and such frames exist in general. Consequently the {\em spin-up}  one-particle  states and the corresponding  vacuum state are well-defined in curved spacetimes.
  \section{One particle states}\label{section-new-action}
 Let $\psi_L$ be the left-handed component of a Dirac field. Suppose that we project $\psi_L$ into its ``spin-up'' component
    \be
    \psi_L^\uparrow:={\hat p}_\uparrow\psi_L,
    \label{psiLup-definition-final}
    \ee
 where ${\hat p}_\uparrow$ is a projection operator whose Lorentz transformation is given by
    \be
    ^\xi\!{\hat p}_\uparrow=U_L(\xi){\hat p}_\uparrow U_L(\xi)^{-1}.
    \label{p-hat-Lorentz-transformation}
    \ee
Therefore, the Lorentz transformation $^\xi\!\psi_L=U_L(\xi)\psi_L$ induces a similar transformation $^\xi\!\psi_L^\uparrow=U_L(\xi)\psi_L^\uparrow$. Consider the operator
   \be
   \cD_L^\uparrow:={\hat p}_\uparrow^\dag \bD_L{\hat p}_\uparrow.
   \label{general-D}
   \ee
Eq.\eqref{Lorentz-D} implies that the Lorentz transformation maps $\bD_L$ to
    \be
    ^\xi\bD_L=U_R(\xi)\bD_LU_L(\xi)^{-1}.
    \ee
 Eq.\eqref{UR-UL*} implies that $U_R(\xi)^\dag=U_L(\xi)^{-1}$, and consequently ${\psi_L^\uparrow}^\dag\cD_L^\uparrow\psi_L^\uparrow$ is invariant under local Lorentz transformations. Therefore, local Lorentz transformation are symmetries of the action
    \be
    \cS[{\psi_L^\uparrow}^\dag,\psi_L^\uparrow]:=i\int d^4x \sqrt{-g}{\psi_L^\uparrow}^\dag\cD_L^\uparrow\psi_L^\uparrow.
    \label{general-action}
    \ee

 The classical field equation reads
     \be
     \cD_L^\uparrow\psi_L^\uparrow=0.
      \label{general-eom}
      \ee
 The path-integral \eqref{1st} and definition \eqref{psiLup-definition-final} imply that  the corresponding Feynman propagator, which we denote by $\bS_F^{\uparrow(L)}(x,x')$, satisfies the identity
    \be
    \bS_F^{\uparrow(L)}(x,x') ={\hat p}_\uparrow\bS_F^{\uparrow(L)}(x,x'){\hat p}_\uparrow^\dag,
    \ee
 and  equations,
    \begin{align}
    \label{general=cond-1}
    &\cD_L^\uparrow\bS_F^{\uparrow(L)}(x,x')=\frac{1}{\sqrt{-g(x')}}\delta^4(x-x'){\hat p}_\uparrow^\dag,\\
    \label{general=cond-2}
    &\bS_F^{\uparrow(L)}(x',x)\circeq-{\bS_F^{\uparrow(L)}(x,x')}^\dag.
    \end{align}

The spin-up component of a left-handed fermion along the $x^3$-direction is separated by
    \be
    \hat p_\uparrow\equiv\frac{1+{\boldsymbol\sigma}_3}{2}.
    \label{general-p+3}
    \ee
In flat spacetime, using Eq.\eqref{general-p+3} and Eq.\eqref{DL} in Eq.\eqref{general-D} we obtain
    \be
    \cD_L^\uparrow=\hat p_\uparrow(\partial_0-\partial_3).
    \ee
Thus the plane-wave solution of the classical field equation \eqref{general-eom} is given by
    \begin{align}
    &\psi_L^\uparrow=e^{-ipx^+}e^{i{\bf q}_\bot\cdot {\bf x}_\bot}\left(\begin{array}{c}1\\0\end{array}\right),&p>0,
    \end{align}
up to a normalization constant, where
    \begin{align}
    \label{general-xpm}
    &x^\pm:=x^0\pm x^3,\\
    &{\bf x}_\bot:=(x^1,x^2),
    \end{align}
and ${\bf q}_\bot\in {\mathbb R}^2$. Following Eqs.\eqref{general=cond-1} and \eqref{general=cond-2} we obtain
     \be
    \bS_F^{\uparrow(L)}=S_F^+(x,x')\delta^2( {\bf x}_\bot- {{\bf x}_\bot'}){\hat p}_\uparrow,
    \label{SF-final}
    \ee
where $S_F^+(x,x')$ is given in Eq.\eqref{SFa}. So, the four dimensional  Feynman propagator is given by the amplitude of left-moving spinors propagating in a  two dimensional subspace.

Now consider a curved spacetime equipped with coordinates $x^\pm$ and ${\bf x}_\bot$  such that
    \be
    g_{--}=0.
    \ee
 Suppose that the Minkowski metric in local frames is given in the light cone gauge \eqref{light-cone-4D}.
  Choose a local frame in which $\partial_{(-)}=(-g)^{-\frac{1}{2}}\partial_-$, i.e.,
    \begin{align}
    \label{general-e--}
    &{e^-}_{(-)}=(-g)^{-\frac{1}{2}},\\
    \label{general-e-mu}
    &{e^\mu}_{(-)}=0,&\mu=+,1,2.
    \end{align}
 The identity
    \begin{align}
    &\eta_{ab}={e^\mu}_{(a)}e_{\mu(b)},&a,b=\pm,1,2,
    \end{align}
 implies that
    \begin{align}
    \label{general-e-+}
    &e_{-(+)}=(-g)^{\frac{1}{2}},\\
    \label{general-e-mu-second}
    &e_{-(a)}=0,&a=-,1,2.
    \end{align}
 These equations also show that for $\mu,\nu=+,1,2$,
    \begin{align}
    \label{general-metric-comp1}
    &g_{-\mu}=(-g)^{\frac{1}{2}} e_{\mu(-)},\\
    \label{general-metric-comp2}
    &g_{\mu\nu}=(-g)^{-\frac{1}{2}}\left(e_{\mu(+)}g_{-\nu}+e_{\nu(+)}g_{-\mu}\right)-\sum_{a=1}^2e_{\mu(a)}e_{\nu(a)}.
    \end{align}

We also  assume that
    \be
    {e^\nu}_{(1)}\nabla_- e_{\nu(2)}=0,
   \label{general-rotation12}
   \ee
 where $\nabla_\mu$ denotes the Levi-Civita connection. This requirement can be satisfied by using  local rotations in the $((1)-(2))$ plane. To see this, start with some tetrad $e'_{\nu(a)}$ satisfying Eqs.\eqref{general-e-+}-\eqref{general-metric-comp2} and define $e_{\nu(\pm)}:=e'_{\nu(\pm)}$ and
    \be
    \left(\begin{array}{c}e_{\nu 1}\\e_{\nu 2}\end{array}\right) :=\left(\begin{array}{cc}\cos \varphi&-\sin\varphi\\ \sin\varphi&\cos\varphi\end{array}\right)\left(\begin{array}{c}e'_{\nu 1}\\e'_{\nu 2}\end{array}\right),
    \ee
 in which $\varphi$ solves the equation
     \be
     \partial_-\varphi={{e'}^\nu}_{(1)}\nabla_-e'_{\nu(2)}.
     \ee
It is easy to verify that $e_{\mu(a)}$ satisfy Eqs.\eqref{general-e-+}-\eqref{general-metric-comp2}  and also Eq.\eqref{general-rotation12}.

Using Eqs.\eqref{general-e--} and \eqref{general-e-mu} in Eq.\eqref{zeta-a} one verifies that
    \be
    {\rm Re}\, \zeta_{(-)}=\frac{1}{2\sqrt{-g(x)}}\partial_\mu\left(\sqrt{-g(x)}{e^\mu}_{(-)}\right)=0.
    \ee
Also, by using  Eqs.\eqref{general-e--}, \eqref{general-e-mu}, \eqref{general-e-mu-second} and \eqref{general-rotation12} in Eq.\eqref{I-mu} one can show that  ${\rm Im}\, \zeta_{(-)}=0$.\footnote{The argument is  similar to footnote \ref{footnote-Imzeta}.} Thus using  Eq.\eqref{general-p+3} in Eq.\eqref{general-D} to separate the spin-up component of $\bD_L$ (Eq.\eqref{DL}) along the third direction, we obtain
    \be
    \cD_L^\uparrow =2\hat p_\uparrow(-g)^{-\frac{1}{2}}\partial_-.
    \label{general-D-final}
    \ee
Eqs.\eqref{general=cond-1} and \eqref{general=cond-2} imply that similarly to the flat spacetime,  the Feynamn propagator is given by Eq.\eqref{SF-final}.

In brief, after using  Eq.\eqref{general-D-final} in Eq.\eqref{general-action} we obtain
    \be
    \cS[{\psi_L^\uparrow}^\dag,\psi_L^\uparrow]:=2i\int d^4x {\psi_L^\uparrow}^\dag\partial_-\psi_L^\uparrow,
    \label{general-action-3rd}
    \ee
and the path integral \eqref{1st} results in Eq.\eqref{SF-final}. Following section \ref{section-upside-down}, the corresponding one-particle states and vacuum state can be postulated similarly to the second quantization in Minkowski spacetime.
\subsection{The Kerr solution}\label{section-Kerr}

The $x^0$ ordering in Eq.\eqref{Int-SF-2D} is a ``time'' ordering only if the vector $\partial_0$ is timelike which is not the case inside an ergosphere. Since
    \be
    {\rm sgn}(x^0-{x'}^0)\delta(x^+-{x'}^+) ={\rm sgn}(x^--{x'}^-)\delta(x^+-{x'}^+),
    \label{switch-ordering}
    \ee
the Feynman propagator  \eqref{Int-SF-2D} can be also understood as an $x^-$ ordered expression
    \be
    S_F^{+}(x,x')=\frac{1}{2}{\rm sgn}(x^--{x'}^-)\delta(x^+-{x'}^+)-\frac{i}{2\pi}\int_0^\infty dp\,\sin\! \left[p(x^+-{x'}^+)\right].
    \label{general-SF-2D}
    \ee

 As an example, consider the Kerr solution whose line element in the Kerr coordinates is given by
    \bea
    ds^2&=&-2dr\left(du^+-a\sin\theta^2d\phi_+\right)\nn\\&-&\rho^{2}d\theta^2-\rho^{-2}\sin\theta^2\left[(r^2+a^2)^2-\Delta a^2\sin\theta^2\right]d{\phi_+}^2\nn\\
    &+&4am\rho^{-2}r\sin\theta^2d\phi_+du_+ +(1-2mr\rho^{-2})d{u_+}^2.
    \label{Kerr-1}
    \eea
where $\rho^2:=r^2+a^2\cos\theta^2$, $\Delta:=r^2+a^2-2mr$, and $m$ and $ma$ are constants representing the mass and the angular momentum as measured from infinity \cite{Hawking-Ellis-book}.  Since $g_{rr}=0$, $\partial_r$ is a null vector and  we can identify $x^-$ with $r$ and choose any suitable function of the other coordinates as $x^+=x^+(u_+,\theta,\phi_+)$. In this way, the Feynman propagator \eqref{general-SF-2D} is  $r$-ordered.

A more familiar description can be obtained  by solving
    \be
    (r^2+a^2)\Delta^{-1}dr=\frac{1}{2}(du_+-du_-),
    \ee
for $r$ and inserting the function $r=r(u_-,u_+)$  in Eq.\eqref{Kerr-1} to obtain the line element in the $(u_\pm,\theta,\phi_+)$ coordinates. In these coordinates $g_{u_-u_-}=0$ and we identify $x^-$ with $u_-$. 
Noting that the Kerr coordinates  in terms of  the   Boyer and Lindquist coordinates $(t,r,\theta,\phi)$  are given by
    \begin{align}
    &du_\pm= dt\pm (r^2+a^2)\Delta^{-1}dr,\\
    &d\phi_+=d\phi+a\Delta^{-1}dr.
    \end{align}
and consequently
    \begin{align}
    &u_-=t- r+2m\ln r+\cO(r^{-1}),
    \end{align}
one verifies that the $u_-$ ordering of the Feynman propagator reproduces the ordinary $t$ ordering via Eq.\eqref{switch-ordering} asymptotically.

\section{Conclusion}\label{section-conclusion}
A spinor field in curved background is defined  by means of local Lorentz transformations. We have shown that in a four dimensional curved background, in general, there exists a spinor field $\psi_L^\uparrow$ which is annihilated by a null vector field $\partial_-$ in a certain frame of reference.
In a  coordinate system given by $x^\pm:=x^0\pm x^3$, and ${\bf x}_\bot:=(x^1,x^2)$ such that the metric component $g_{--}=0$, this frame of reference is identified by the following conditions on the tetrad ${e^\mu}_{(a)}$,
     \begin{align}
     \label{concl-frame-1}
     &\partial_{(-)}=(-g)^{-\frac{1}{2}}\partial_-,\\
     \label{concl-frame-2}
     &{e^\nu}_{(1)}\nabla_- e_{\nu(2)}=0,
     \end{align}
where $\nabla_\mu$ denotes the Levi-Civita connection, $g$ is the determinant of the spacetime metric and  $\partial_{(a)}:={e^\mu}_{(a)}\partial_\mu$, and
    \be
     \psi_L^\uparrow:=\frac{1+{\boldsymbol\sigma}^3}{2}\psi_L,
    \ee
in which, ${\boldsymbol\sigma}^3$ is the third Pauli matrix and $\psi_L$ is a left-handed massless Dirac field.

 The corresponding Feynman propagator is given by
    \be
    S_F^{\uparrow(L)}=S_F^+(x,x')\delta^2( {\bf x}_\bot- {{\bf x}_\bot'}),
    \label{concl-1}
    \ee
 in which $S_F^+(x,x')$  denotes the Feynman propagator obtained by means of the second quantization of a left-moving massless Dirac field in  two dimensional Minkowski spacetime. Therefore, $S_F^{\uparrow(L)}$ can be interpreted in terms of propagating one-particle states confined to a two dimensional Minkowski spacetime equipped with coordinates $x^\pm$, and the corresponding vacuum state is well-defined similarly to the second quantization in  Minkowski spacetime. In the flat spacetime limit, $\psi_L^\uparrow$ is reminiscent of  the spin-up component of a left-handed massless Dirac field travelling along the $x^3$-axis.

This line of thought is motivated by  an observation in two dimensions. As we have argued in detail, in a two dimensional curved background there exists, in general, a local frame in which   the left-moving massless Dirac field is annihilated by a null vector field $\partial_-$, and consequently, the corresponding Feynman propagator equals $S_F^+(x,x')$.  Therefore, the Feynman propagator can be interpreted in terms of propagating one-particle states similarly to the second quantization in Minkowski spacetime. In such local frames, the curvature effect is totally transmitted  to the right-moving sector. Consequently,  the interpretation of the corresponding Feynman propagator in terms of propagating right-moving modes requires an $x$-dependent normalization of the one-particle states.

 In four dimensions,  the chirality is reversed by CPT transformation \cite{Alvaerz-Gaume-Witten}, hence,  both of the left-handed and the right-handed components of the massless Dirac field  are equally affected by the spacetime curvature. So we have focused on the left-handed sector and separated its spin-up and spin-down components covariantly  with respect to the local Lorentz transformations. We have introduced an action  for the spin-up component which enjoys diffeomorphism invariance and local Lorentz transformation. It is given by
    \be
    \cS[{\psi_L^\uparrow}^\dag,\psi_L^\uparrow]:=i\int d^4x \sqrt{-g}{\psi_L^\uparrow}^\dag\bD_L\psi_L^\uparrow,
    \ee
 where $\bD_L$ is the Dirac operator for massless fermions in the left-handed sector, and  $\psi_L^\uparrow$ is the spin-up component of the left-handed Dirac field $\psi_L$
    \be
    \psi_L^\uparrow :={\hat p}_\uparrow\psi_L,
    \ee
 in which ${\hat p}_\uparrow$ is the corresponding projection operator. We have supposed that $\psi_L^\uparrow$ is in the spin $\frac{1}{2}$ representation of  the local Lorentz transformations similarly to $\psi_L$. This can be done by considering ${\hat p}_\uparrow$ as a tensor field. Explicitly, if $U_L(\Lambda)$ denotes the operator corresponding to a local Lorentz transformation  $\Lambda$  in the left-handed sector such that $\psi_L\to U_L(\Lambda)\psi_L$, we require that
    \be
    {\hat p}_\uparrow\to U_L(\Lambda){\hat p}_\uparrow U_L(\Lambda)^{-1},
    \ee
accordingly.
 In the frame of reference given by Eqs.\eqref{concl-frame-1} and \eqref{concl-frame-2},  the projection operator ${\hat p}_\uparrow$   equals $\frac{1+{\boldsymbol\sigma}^3}{2}$ and
    \be
    \cS[{\psi_L^\uparrow}^\dag,\psi_L^\uparrow]:=2i\int d^4x {\psi_L^\uparrow}^\dag\partial_-\psi_L^\uparrow.
    \ee
 Therefore, the classical field equation implies that $\psi_L^\uparrow$ is annihilated by the null vector $\partial_-$ and the Feynman propagator is given by Eq.\eqref{concl-1}.
 Consequently, the notion of fermionic one-particle states and the corresponding vacuum state is well-defined in four dimensional (non-stationary) curved backgrounds. Such particles   travel without being scattered by the background geometry.


\ed